\documentclass[aip, apl, reprint]{revtex4-1}
\usepackage{graphicx}
\usepackage{dcolumn}
\usepackage{bm}
\usepackage[utf8]{inputenc}
\usepackage[T1]{fontenc}
\usepackage{mathptmx}
\usepackage{etoolbox}

\makeatletter
\def\@email#1#2{%
 \endgroup
 \patchcmd{\titleblock@produce}
  {\frontmatter@RRAPformat}
  {\frontmatter@RRAPformat{\produce@RRAP{*#1\href{mailto:#2}{#2}}}\frontmatter@RRAPformat}
  {}{}
}%
\makeatother
\begin{document}

\title{Using \textit{k}-means to sort spectra: electronic order mapping from scanning tunneling spectroscopy measurements}

\author{V. King}
\email{vanessa.king@chem.ubc.ca}
\affiliation{Stewart Blusson Quantum Matter Institute, University of British Columbia, Vancouver, V6T 1Z4, Canada}
\affiliation{Department of Chemistry, University of British Columbia, Vancouver, V6T 1Z1, Canada}

\author{Seokhwan Choi}
\affiliation{Stewart Blusson Quantum Matter Institute, University of British Columbia, Vancouver, V6T 1Z4, Canada}

\author{Dong Chen}
\affiliation{Stewart Blusson Quantum Matter Institute, University of British Columbia, Vancouver, V6T 1Z4, Canada}
\affiliation{Department of Physics and Astronomy, University of British Columbia, Vancouver, V6T 1Z1, Canada}

\author{Brandon Stuart}
\affiliation{Stewart Blusson Quantum Matter Institute, University of British Columbia, Vancouver, V6T 1Z4, Canada}
\affiliation{Department of Physics and Astronomy, University of British Columbia, Vancouver, V6T 1Z1, Canada}

\author{Jisun Kim}
\affiliation{Stewart Blusson Quantum Matter Institute, University of British Columbia, Vancouver, V6T 1Z4, Canada}

\author{Mohamed Oudah}
\affiliation{Stewart Blusson Quantum Matter Institute, University of British Columbia, Vancouver, V6T 1Z4, Canada}
\affiliation{Department of Physics and Astronomy, University of British Columbia, Vancouver, V6T 1Z1, Canada}

\author{Jimin Kim}
\affiliation{Center for Artificial Low Dimensional Electronic Systems, Institute for Basic Science, Pohang, 37673, South Korea}
\affiliation{Department of Physics, Pohang University of Science and Technology,
Pohang, 37673, South Korea}

\author{B. J. Kim}
\affiliation{Center for Artificial Low Dimensional Electronic Systems, Institute for Basic Science, Pohang, 37673, South Korea}
\affiliation{Department of Physics, Pohang University of Science and Technology,
Pohang, 37673, South Korea}

\author{D. A. Bonn}
\affiliation{Stewart Blusson Quantum Matter Institute, University of British Columbia, Vancouver, V6T 1Z4, Canada}
\affiliation{Department of Physics and Astronomy, University of British Columbia, Vancouver, V6T 1Z1, Canada}

\author{S. A. Burke}
\affiliation{Stewart Blusson Quantum Matter Institute, University of British Columbia, Vancouver, V6T 1Z4, Canada}
\affiliation{Department of Chemistry, University of British Columbia, Vancouver, V6T 1Z1, Canada}
\affiliation{Department of Physics and Astronomy, University of British Columbia, Vancouver, V6T 1Z1, Canada}

\date{\today}

\begin{abstract}
Hyperspectral imaging techniques have a unique ability to probe the inhomogeneity of material properties whether driven by compositional variation or other forms of phase segregation. In the doped cuprates, iridates, and related materials, scanning tunneling microscopy/spectroscopy (STM/STS) measurements have found the emergence of pseudogap `puddles' from the macroscopically Mott insulating phase with increased doping. However, categorizing this hyperspectral data by electronic order is not trivial, and has often been done with ad hoc methods. In this paper we demonstrate the utility of \textit{k}-means, a simple and easy-to-use unsupervised clustering method, as a tool for classifying heterogeneous scanning tunneling spectroscopy data by electronic order for Rh-doped Sr$_2$IrO$_{4}$, a cuprate-like material. Applied to STM data acquired within the Mott phase, \textit{k}-means successfully identified areas of Mott order and of pseudogap order. The unsupervised nature of \textit{k}-means limits avenues for bias, and provides clustered spectral shapes without \textit{a priori} knowledge of the physics. Additionally, we demonstrate successful use of \textit{k}-means as a preprocessing tool to constrain phenomenological function fitting. Clustering the data allows us to reduce the fitting parameter space, limiting over-fitting. We suggest \textit{k}-means as a fast, simple model for processing hyperspectral data on materials of mixed electronic order.
\end{abstract}
\maketitle 

Understanding heterogeneous electronic behaviour in quantum materials is critical to our knowledge of the physics governing electronic order. Scanning tunneling microscopy and spectroscopy (STM/STS) are ideal methods to probe heterogeneous electronic order, including in doping variations in 2D and Dirac materials\cite{Zhang2009,Beidenkopf2011,Shin2016}, distinct surface terminations\cite{Zou2023}, and electronic phase segregation like seen in some correlated systems\cite{Chang1992,Pan2001,Lang2002,Chang2007,Singh2013}. In this work, we explore the use of k-means clustering to label spatially resolved STS, though the analysis methods we will discuss need not be limited to STM/STS\cite{Ziatdinov2019,Borodinov2020,Melton2020,Kaming2021,Sun2021}.
The spatially heterogeneous electronic structure of cuprates and similar correlated electron systems are a significant area of interest. This class of materials undergoes a Mott insulator to metal transition upon sufficient doping. The phase transition regime is characterized by phase segregation of regions with distinct electronic order, featuring `puddles' of pseudogap order that appear amongst the Mott insulating phase\cite{Kohsaka2004,Alldredge2008,Kim2010,Maksymovych2022,Kim2016,Battisti2017,Jiang2019} providing an ideal and physically meaningful test case for applying clustering tools.

STM and STS are powerful tools for characterizing materials with spatially heterogeneous electronic structure, as these methods are sensitive to the local density of states (LDOS) as a function of energy.
Spectroscopic imaging -- where STS is acquired at each pixel -- allows us to both spatially and spectrally resolve the electronic structure. However, deducing electronic order from these spectra is challenging; it often requires fitting highly parameterized physical or empirical models to the point spectra, or using ad-hoc methods like comparing selected energy regions to generate the necessary contrast. The challenges associated with this task are in some ways surprising given the differences in the spectra are often easily discernible by eye. Earlier analyses done on cuprates and cuprate-like materials looked at an energy slice where the Mott gap and pseudogap orders were most distinct and choose a LDOS threshold to classify electronic order.\cite{Zhang2009,Beidenkopf2011} This method reduces the energy dimension of the data set, thereby losing most of the information on the spectral shape, and risks misclassification of noisy data. More comprehensive approaches have used a fitting function based on the Dynes formula to quantify the pseudogap.\cite{Alldredge2008,Kohsaka2012,Torre2015,Kim2016} Most recently, such hyperspectral data was fit with a phenomenological term for both the Mott gap and the pseudogap, to extract both gap parameters\cite{Battisti2017}, however such multicomponent fits risk being poorly constrained due to the large number of parameters needed to describe data where there is a coexistence of multiple electronic orders.

\begin{figure}[!ht]
   \includegraphics[width=8.5cm]{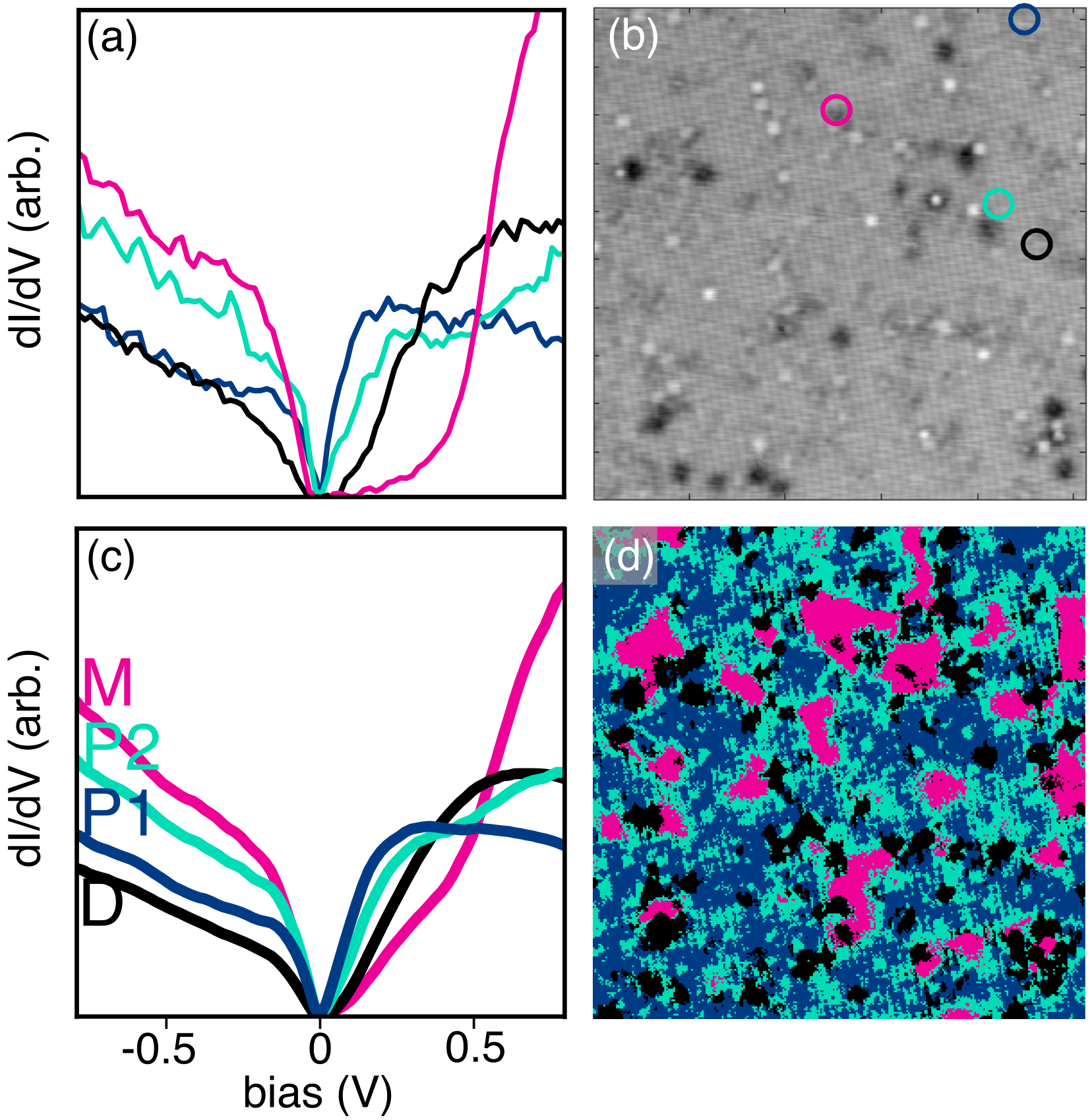}
    \caption{ (a) Example spectra picked out by eye from (b) a 24 nm x 24 nm STM topography image of 5.3\% Rh-doped Sr$_2$IrO$_{4}$, taken at 800 mV and 200 pA. Positions of the spectra from (a) are circled. (c) -- (d) results from \textit{k}-means applied to the same grid map for 4 clusters. (c) centroids of the 4 clusters, named according to their spectral shape. (d) spatial distribution of the 4 clusters.}\label{fig:1}
\end{figure}

With the increasing accessibility of machine learning tools to non-specialists, there has been a growing interest in leveraging machine learning approaches to analyze hyperspectral data.\cite{Rickman2019,Gordon2020Review} However, the low quantity and high variance of data has made it challenging to apply supervised algorithms to hyperspectral techniques such as STM or atomic force microscopy (AFM). This can be partially resolved by padding the training sets with artificial data generated by an \textit{a priori} physical model, however, this can introduce additional sources of bias.\cite{Ziatdinov2019,Zhang2019,Gordon2020,Alldritt2020,Wang2020,Roccapriore2021} Unsupervised models have the benefit of not requiring training data, though the trade-off is that they are difficult to tune; if an unsupervised model doesn't perform as desired, there likely isn't much that can be done. Nonetheless, unsupervised models have successfully been applied to hyperspectral data for example: the use of principal component analysis or variational autoencoders for dimensionality reduction, unsupervised Bayesian linear unmixing for regression, and \textit{k}-means for clustering.\cite{Belianinov2014,Ziatdinov2016,Kannan2018,Ito2018,Wahl2020,Melton2020,Roccapriore2021,Kaming2021,Sun2021,Ziatdinov2022,Maksymovych2022,Kalinin2023,Zou2023}

Here, we apply \textit{k}-means -- a widely used and highly accessible machine learning clustering algorithm -- to categorize qualitatively distinct spectra from an STS imaging grid and distinguish the Mott gap and pseudogap regions of a spatially heterogeneous sample. \textit{k}-means is unsupervised, which allows us to avoid defining expected electronic states, enabling the discovery of unexpected electronic behaviour. While clustering does not generate parameter estimates, it is useful as a preliminary analysis step that reduces the number of free variables for subsequent function fitting of the clustered data. Indeed, unsupervised clustering algorithms such as \textit{k}-means have already been shown to produce reliable clusters for use in ensuing analysis.\cite{Ziatdinov2016,Li2017,Ziatdinov2019,Farley2020,Borodinov2020}

In this letter, we used Rh-doped strontium iridate samples in a doping range spanning the bulk Mott-pseudogap phase transition as a test case. Doped iridates exhibit a heterogeneous electronic structure with segregation of pseudogap phase and Mott insulating phase\cite{Torre2015,Kim2016,Battisti2017} also characteristic of the cuprates\cite{Chang1992,Pan2001,Lang2002,Kohsaka2004,Alldredge2008,Kohsaka2012} and other similar materials\cite{Kim2010}.
We will use \textit{k}-means as a tool for identifying and mapping the electronic order across STM/STS grid maps of each sample. Then, we will use the \textit{k}-means results to inform a constrained function fitting analysis for gap size estimation.

Here we present the \textit{k}-means analysis of three STM/STS spectroscopic imaging measurements on doped iridate samples. The first is a 5.3\% Rh-doped Sr$_2$IrO$_{4}$ sample measured at 4.5 K, with the STM topography shown in Figure \ref{fig:1} (b). As this is within the bulk Mott insulating phase\cite{Qi2012,Zhao2016}, we expect this to show a mixed Mott-psuedogap order. Indeed, the example spectra shown in Figure \ref{fig:1} (a) show spatially separated regions with spectra characteristic of a Mott insulator and pseudogap. We now look to \textit{k}-means to map the different regions of this heterogeneous electronic behaviour.

The \textit{k}-means algorithm requires the user to specify the number of clusters, $k$, and then generates a random `centroid' to represent each cluster, here a spectral shape. The algorithm first assigns each data point (in our case each point spectrum at a specific x,y coordinate) to its nearest centroid by Euclidean distance. For each cluster, $\boldsymbol{S}_j$, it computes a new centroid, $\boldsymbol{\mu}_j$, by taking the mean of all the data points, $\boldsymbol{s}_i$, assigned to that cluster. These steps, as illustrated by a toy model shown in supplemental Figure 1 are repeated until the change in the within-cluster variance between iterations falls below a specified threshold. In effect, this algorithm minimizes the within-cluster sum of squares (WCSS), or `inertia', as shown in equation \ref{eq:inertia}.\cite{Belianinov2014} 

\begin{equation}\label{eq:inertia}
    \mbox{arg min }\sum^{k}_{j=1}\sum_{\boldsymbol{s}_i\in\boldsymbol{S}_j} \|\boldsymbol{s}_i - \boldsymbol{\mu}_j\|^2.
\end{equation}

One of the largest challenges of using \textit{k}-means is in deciding on the appropriate number of clusters. The number of clusters \textit{k} can be inferred from an \textit{a priori} physical model, but this can limit discovery of behaviour that is outside of expectations.\cite{Sun2021} Qualitative differences in the \textit{k}-means centroids are often used as evidence that each division of clusters is meaningful.\cite{Ziatdinov2019,Borodinov2020,Maksymovych2022} If these methods aren't applicable, then a plot of minimized inertia values versus $k$ can be visually examined for a distinct kink or `elbow'.\cite{Kaming2021} Alternatively, one can search for an `elbow' in other metrics such as, the Calinsky-Harabasz score, which characterizes the WCSS over the between-cluster sum of squares (BCSS).\cite{scikit,Kaming2021} However, there is no guarantee that the \textit{k}-means results for a given data set will contain a visually obvious elbow for any of these parameters. Additional methods of choosing $k$ include performing hierarchical clustering on the data and examining the shape of the resulting dendogram.\cite{Belianinov2014,Ziatdinov2016,Sun2021} There are numerous other analytical methods that try to more quantitatively determine the appropriate $k$, but each depends on some \textit{a priori} knowledge of the data set's statistical qualities. As a result, determining the optimal $k$ remains somewhat of an art, and a trial-and-error approach may be required.

For our data set, we implemented \textit{k}-means using the pre-built \text{k}-means function in the scikit-learn library.\cite{scikit} This implementation of \textit{k}-means runs on a data set consisting of 75,625 spectra, each with 81 points in energy in well under 30 seconds on a laptop computer. We initialized the algorithm using the `k-means++' method, which entails choosing initial centroids that are more distant from each other and decreases the number of iterations required to reach convergence. Alternative methods for setting the initial centroids include randomly selecting point spectra from the data or providing spectra that represent the expected electronic states (shown in Figure \ref{fig:1} (a)). With enough iterations, all three of these approaches converged to near-identical results for our data, as shown in supplemental Figure 3. The scikit-learn implementation of \textit{k}-means allows the user to define the $\Delta$WCSS threshold for convergence, and the maximum number of iterations to try to reach that convergence. It also allows for the entire \textit{k}-means algorithm to be repeated multiple times, with different initial centroids, to increase the likelihood that it finds a global minimum. Again, we found that choosing any reasonable values of these parameters did not significantly impact the clustering of our data. 

\begin{figure}[!h]
    \includegraphics[width=8.5cm]{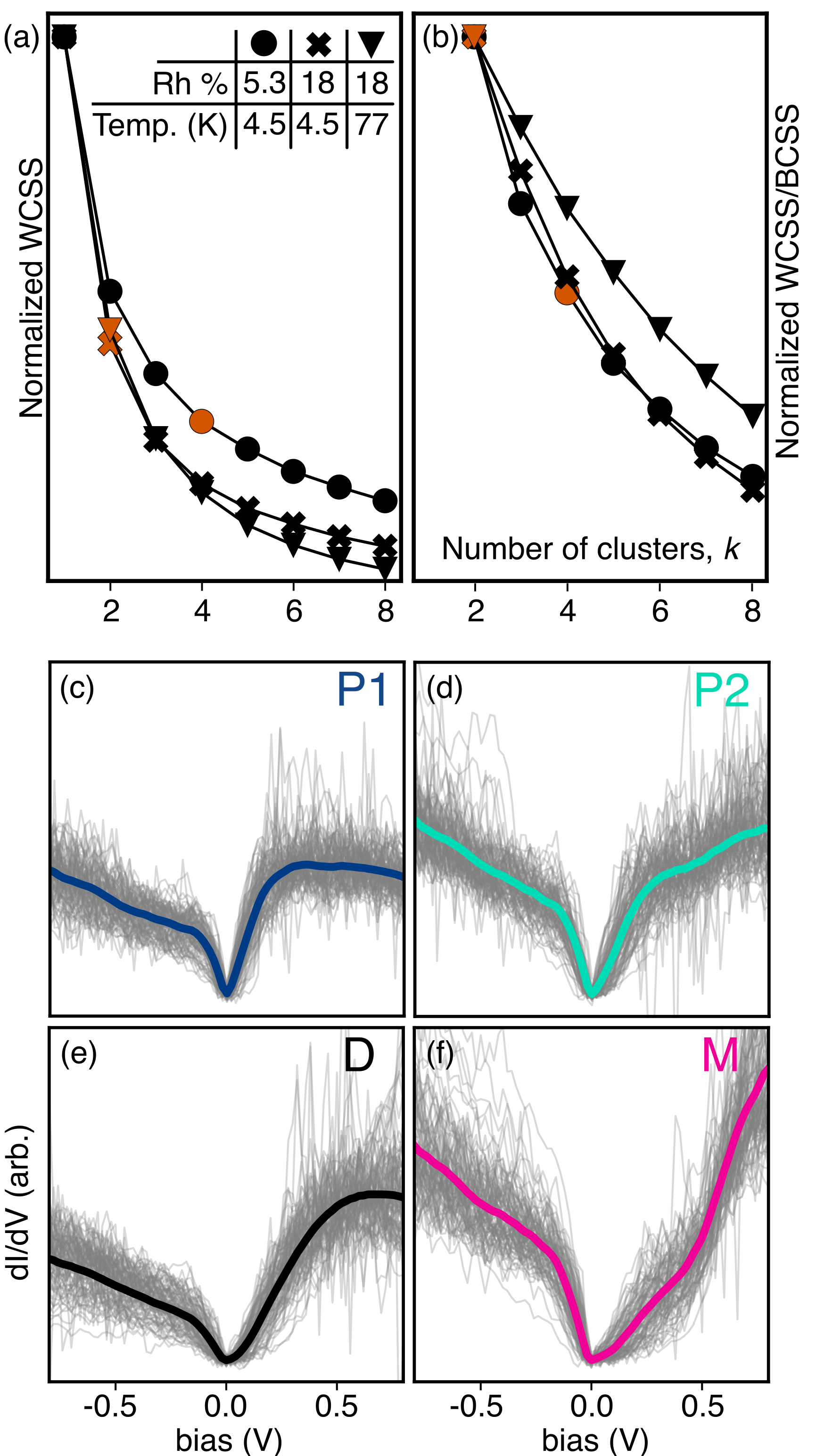}
    \caption{ (a) -- (b) Selected error metrics for the \textit{k}-means results on Rh-doped Sr$_2$IrO$_{4}$ samples for different values of $k$. The chosen number of clusters for each dataset highlighted red. (a)
    Normalized within-cluster sum-of-squares (WCSS). (b) Normalized WCSS over the between-cluster sum-of-squares (BCSS).
    (c) -- (f) centroids and randomly sampled spectra belonging to the respective cluster.}\label{fig:2}
\end{figure}

When applying \textit{k}-means to data from the 5.3\% Rh-doped Sr$_2$IrO$_{4}$ sample at 4.5 K where the sample is macroscopically a Mott insulator but locally heterogeneous, we did not see an obvious `elbow' in the WCSS or WCSS/BCSS scores, as shown in Figures \ref{fig:2} (a) and (b). Instead we based our choice of $k$ on physical intuition and a qualitative assessment of centroid shapes. Given that we see a mix of pseudogap and Mott gap order (Figure \ref{fig:1} (a)), and we see several defects in the topography of Figure \ref{fig:1} (b) which also have distinct spectra, we deduced that our sample should have at least 3 spectrally distinct clusters. We started with $k=3$ and incremented until the new centroids stopped showing qualitatively distinct functional shapes, giving us a value of $k=4$. 

These 4 clusters consist of two pseudogap clusters, one Mott cluster, and a defect cluster. This was determined by comparing the centroid shapes in Figure \ref{fig:1} (c) to the example spectra in Figure \ref{fig:1} (a). The two pseudogap clusters appear to generally follow the lattice seen in the topography image in Figure \ref{fig:1} (b). Additionally, the pseudogap 2 (P2) centroid has roughly symmetric gap shoulders, while the pseudogap 1 (P1) centroid has a notably lower negative bias gap shoulder. This qualitative difference suggests that the two clusters have distinct spectral signatures, and are not just gradations of a single pseudogap order. A visual examination of Figures \ref{fig:2} (c) -- (f) shows that the spectra are appropriately clustered, and thus it is reasonable to use these clusters to delineate distinct spectra.

The other two measurements analyzed are of an 18\% Rh-doped Sr$_2$IrO$_{4}$ sample measured at 4.5 K and 77 K, shown in supplemental Figures 4 -- 7. It is thought that the 18\% Rh sample is in the hidden order phase at both temperatures\cite{Zhao2016}, but it is not clear if there would be any remaining Mott insulating phase. For both of the 18\% Rh measurements, we found that no value of $k$ resulted in a centroid with a Mott gap shape or defect spectra shape (supplemental Figures 5 and 7), and manual examination of all the spectra for these measurements found no Mott gap areas. Thus, $k=2$ captured all of the qualitatively different spectral groups: two pseudogap clusters, demonstrating that this approach with a physically motivated starting point and seeking qualitatively distinct clusters by incrementing $k$ can be used to explore the physical characteristics appearing in this type of data set.

To contrast the \textit{k}-means approach, we also applied a recent analysis used for a similar doped iridate sample relying on fits with phenomenological terms for both the Mott gap and pseudogap to extract gap parameters, as shown in equation \ref{eq:BMP}.\cite{Battisti2017}

\begin{equation}\label{eq:BMP}
    DOS(V) = B(V)+ M(V) + P(V).
\end{equation}
The first term, $B(V)$ accounts for any background $\frac{d I}{d V}$ intensity not attributable to the low-energy electronic structure (Mott gap and pseudogap). The definition of the background term is often tailored to the dataset to improve the overall fit quality of equation \ref{eq:BMP}. We have defined our background term using equation \ref{eq:B}, to account for a background density of states and influence of the transmission function at larger energy,
\begin{equation}\label{eq:B}
    B(V) = c_1 V^2 + c_2 V + c_3.
\end{equation}
The last two terms of equation \ref{eq:BMP} are phenomenological fits for a Mott gap, $M(V)$ and pseudogap, $P(V)$, respectively:
\begin{equation}\label{eq:M}
    M(V) = c_M\left(1- \left|\frac{1}{1+e^{(V-e_0)/\omega}} - \frac{1}{1+e^{(V+e_0 - \Delta_M)/\omega}} \right|\right),
\end{equation}
\begin{equation}\label{eq:P}
    P(V) = c_P\left|\frac{|V| + i\alpha\sqrt{|V|}}{\sqrt{\left(|V|+i\alpha\sqrt{|V|} \right)^2 - \Delta_P^2}} \right|,
\end{equation}

where $c_M$, $c_P$, and $c_1$ through $c_3$ are coefficients weighting each term, $e_0$ is the energy where the upper Hubbard band is confined relative to the chemical potential, $\omega$ is an arbitrary parameter that broadens the Mott gap edges, $\Delta_M$ and $\Delta_P$ are the sizes of the Mott gap and pseudogap respectively, and $\alpha$ is the effective scattering rate. This leads to a total of 10 free parameters in the fit in order to describe all regions of the sample by one function. However, we do not expect that a given point in space physically consists of a combination of electronic orders: ideally the coefficients $c_M$ and $c_P$ should drop to zero in the pseudogap and Mott regions. Consequently, fitting a spectrum to both the Mott gap and pseudogap terms (as is done in equation \ref{eq:BMP}) may not yield accurate values and risks over-fitting. 

\begin{figure}[!h]
    \includegraphics[width=8.5cm]{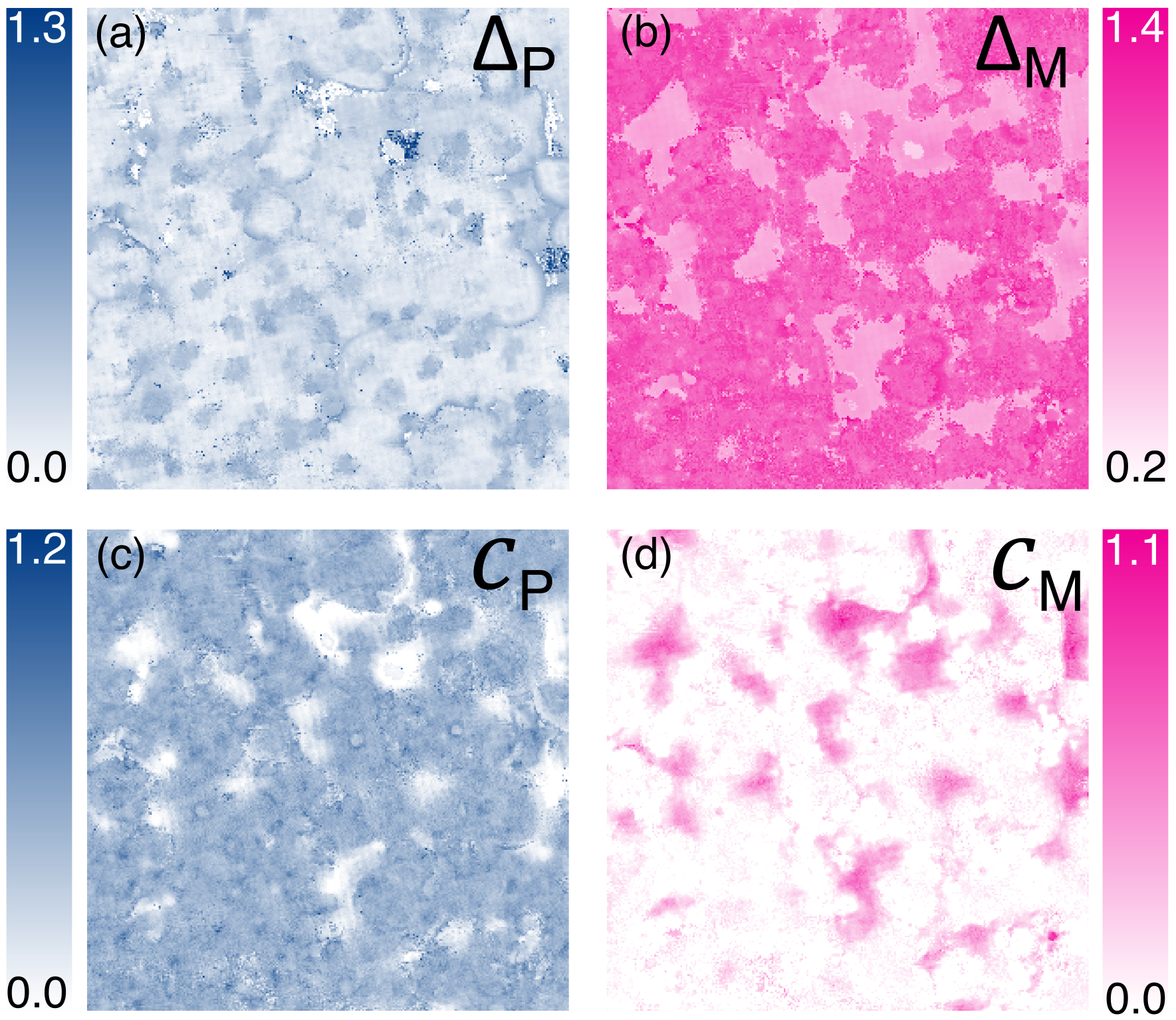}
    \caption{ Typical (a)$\Delta_P$ and (b) $\Delta_M$ maps calculated by fitting all spectra to equation \ref{eq:BMP}. Maps of the leading coefficient of the (c) pseudogap ($c_P$) and (d) Mott ($c_M$) terms of equation \ref{eq:BMP} created from the 5.3\% Rh-doped Sr$_2$IrO$_{4}$ grid map.}\label{fig:3}
\end{figure} 

Figure \ref{fig:3} shows the results of such an analysis on the 5.3\% Rh-doped Sr$_2$IrO$_{4}$ data set. Figures \ref{fig:3} (a) and (b) are representative of the typical `gap maps' produced for analysis of inhomogeneous cuprate and cuprate-like materials.\cite{Alldredge2008,Kohsaka2012,Torre2015,Kim2016,Battisti2017} These gap maps, alone, are not as effective at identifying the spatial distribution of electronic order. Figures \ref{fig:3} (c) and (d) show a map of the pseudogap term ($c_P$) and Mott term ($c_M$), respectively. With the insight of the \textit{k}-means results, we can see that the `gap maps' are only providing relevant gap size information in areas where the coefficients shown in Figures \ref{fig:3} (c) and (d) are greater. Even if one was to try to classify the spatial distribution of order using a combined `gap map' and so-called `coefficient map', one would need to manually pick a coefficient threshold value, introducing an additional source of bias. Unsupervised machine learning algorithms such as \textit{k}-means allow us to avoid many of these analysis steps and user decisions.

\begin{figure*}[ht!]
    \includegraphics[width=17cm]{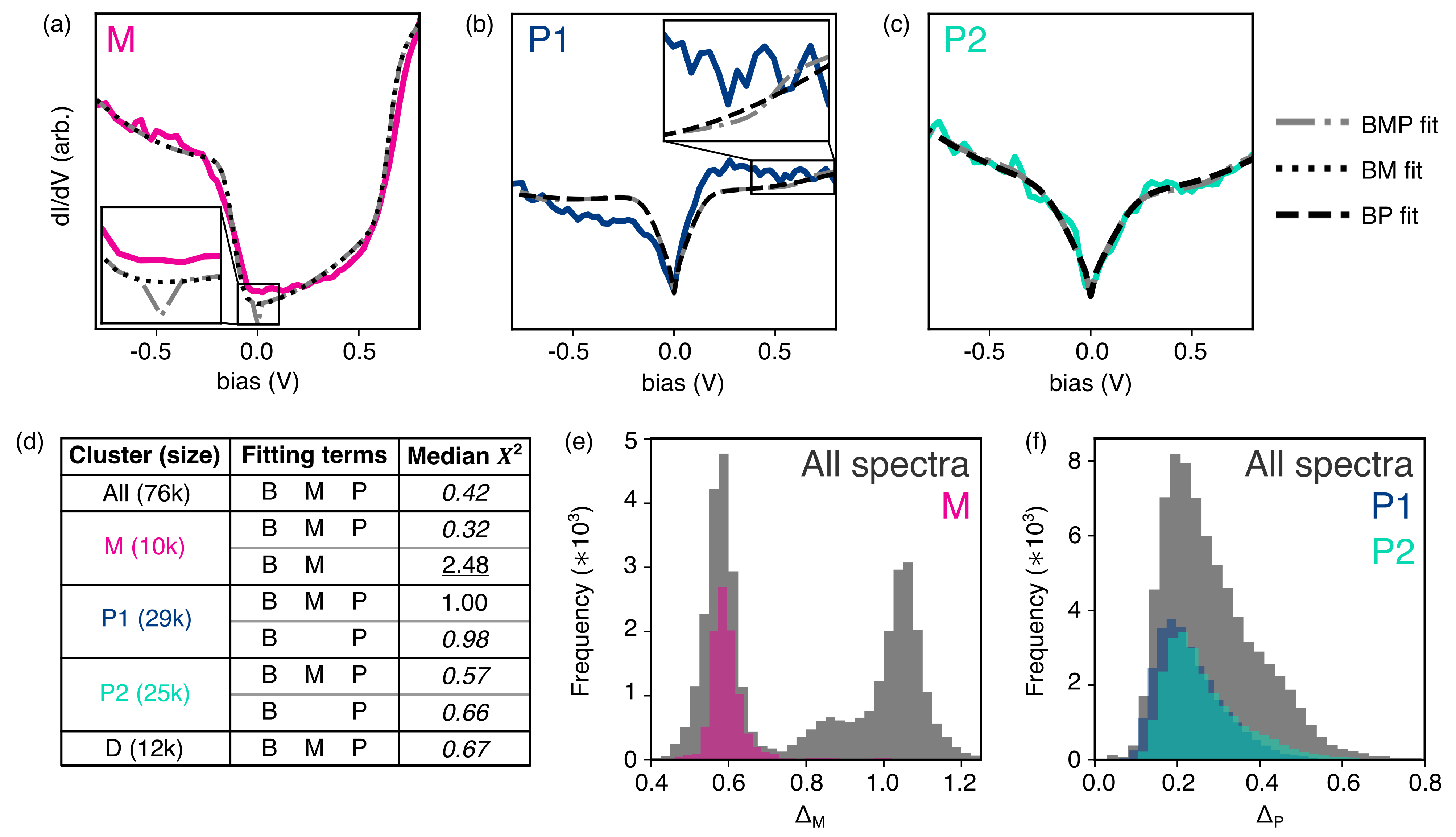}
    \caption{ (a) -- (c) fits of two or three terms to the example spectra from Figure \ref{fig:1} (a). (d) median $\chi^2$ values
    for 2 vs 3 fitting terms applied to different clusters of spectra.
    (e) -- (f) distribution of $\Delta_M$ and $\Delta_P$ when fit with all fitting terms to different clusters of spectra.}\label{fig:4}
\end{figure*}

If in addition to classification we require parameter estimates, we can instead use the successfully clustered data to fit more constrained models containing the minimum parameters. For the Mott cluster and pseudogap clusters we can remove the pseudogap term (equation \ref{eq:P}) and the Mott term (equation \ref{eq:M}), respectively, from the fitting function. This helps to reduce over-fitting; as shown in Figures \ref{fig:4} (a) -- (c), fits with both Mott gap and pseudogap terms add erroneous features not reflected by the data or simply don't improve the fit. Figure \ref{fig:4} (d) shows that nearly all fits exhibit over-fitting (median $\chi^2 < 1$) with the Mott-clustered spectra fit with the full equation \ref{eq:BMP} being the most egregious. Notably, removing the pseudogap term (equation \ref{eq:P}) from the fits to spectra in the Mott cluster increases the $\chi^2$ value, implying it is no longer over-fitting the data. In addition to allowing for more constrained fitting functions, the centroids have the potential to reveal electronic behaviour that isn't modelled by phenomenological equations. In Figure \ref{fig:4} (b), we see that the qualitative shape of the sample spectrum of cluster P2 is not well captured by equation \ref{eq:P}, given the asymmetry of the gap shoulders. This example demonstrates the utility of \textit{k}-means in revealing unexpected electronic features.

Figures \ref{fig:4} (e) and (f) show how the fitted $\Delta_M$ and $\Delta_P$ values for the clustered data are each a distinct subset of the total distribution. In particular, the clustering results indicate that only one of the peaks of the $\Delta_M$ distribution for all spectra is actually associated with a Mott gap. Thus a Mott gap or pseudogap map over all spectra may give a misleading impression of the distribution of such physically meaningful parameters, as seen in Figures \ref{fig:3} (a) and (b).

\textit{k}-means has enabled us to sort and categorize the spatially resolved tunneling spectra of a mixed electronic order sample with minimal user input. The successful test case used exhibited a high degree of electronic heterogeneity with qualitatively different spectral shapes. Other use cases with less distinct spectra, or more sparsely populated regions may be less well suited to \textit{k}-means.  We expect \textit{k}-means to aid in the discovery of unanticipated electronic behaviour that is not well represented by phenomenological fitting functions. For well established electronic orders, \textit{k}-means clustering is a promising preliminary method that enables use of more constrained models.

We encourage others examining electronically heterogeneous hyperspectral data to incorporate \textit{k}-means into their analyses, both because of its analytical merits and its ease of use. We would recommend inputting  hyperspectral data as a set of independent spectra into any library implementation of \textit{k}-means. Starting with an estimate of \textit{k} based on physical knowledge or observation, visual inspection of the qualitative behaviour of the centroids, or metrics like inertia, can be used to refine an appropriate \textit{k} value for your system. Where applicable, more constrained physical models can be fit to the clustered spectra to extract physically meaningful parameters of the system.

\vspace{4mm}
See the supplemental materials for crystal growth and measurement details, further visualization of \textit{k}-means functionality, and \textit{k}-means analysis of additional datasets.\\

\begin{acknowledgments}
The authors would like to thank R. Krems for helpful discussions. This research was undertaken thanks in part to funding from the Max Planck-UBC-UTokyo Centre for Quantum Materials (M.O.), and the Canada First Excellence Research Fund in Quantum Materials and Future Technologies Program. V.K. acknowledges the support of the British Columbia Graduate Scholarship. This work was also supported by the NSERC Discovery Grants and Research Tools and Instruments programs, the Canadian Foundation for Innovation and British Columbia Knowledge Development Fund, the Canada Research Chairs program (S.A.B.).
\end{acknowledgments}

\section*{Author Declarations}
\subsection*{Conflict of Interest}
The authors have no conflicts to disclose.

\subsection*{Author Contributions}
\textbf{V. King}: Conceptualization (equal), Data curation (lead), Formal analysis (lead), Investigation (lead), Methodology (lead), Software (lead), Writing -- original draft (lead), Writing -- review \& editing (equal). 
\textbf{Seokhwan Choi}: Investigation (supporting).
\textbf{Dong Chen}: Investigation (supporting).
\textbf{Brandon Stuart}: Investigation (supporting).
\textbf{Jisun Kim}: Investigation (supporting), Writing -- review \& editing (supporting).
\textbf{Mohamed Oudah}: Resources (equal).
\textbf{Jimin Kim}: Resources (equal).
\textbf{B. J. Kim}: Resources (equal).
\textbf{D. A. Bonn}: Conceptualization (equal), Funding acquisition (equal), Supervision (equal), Writing -- review \& editing (equal).
\textbf{S. A. Burke}: Conceptualization (equal), Funding acquisition (equal), Supervision (equal), Writing -- review \& editing (equal).

\section*{Data Availability Statement}
The data that support the findings of this study are openly available in OSF at http://doi.org/10.17605/OSF.IO/XA35N, reference number \cite{OSF_repository} along with a sample script for the basic application of \textit{k}-means on hyperspectral data.

\section*{References}
\bibliography{k-means}

\end{document}


\title{Supplemental Materials}
\date{}
\maketitle 

\section*{Crystal growth and sample preparation}\label{sup:sample}
Rh-doped Sr$_2$IrO$_4$ single crystals were grown using a self-flux method. IrO$_2$ (99.99\%), SrCO$_3$ (99.994\%), and SrCl$_2\cdot6$H$_2$O (99.9965\%) powders were thoroughly mixed and ground in an agate mortar, and placed in a platinum crucible covered with a platinum lid. For Rh doping, we deliberately added extra RhO$_2$ (75.2-77.4\%) powders. All powders were bought from Thermo Fisher Scientific. Crucibles were enclosed by an outer alumina crucible and heated in a programmable box furnace in air. We used the same heating sequence as previously reported\cite{GrowthRecipe}. Square-shaped single crystals were extracted from the residual flux in the crucible by rinsing out with distilled water. The crystals were characterized using powder X-ray diffraction (XRD) with Cu-K$\alpha1$ radiation, and the ratio of elements were confirmed using a scanning electron microscope equipped with an energy dispersive X-ray spectrometer. Samples were cleaved \textit{in situ} before transfer into the STM chamber.

\section*{STM/STS Measurements}\label{sup:STM}
All SPM data was taken in an ultra-high vacuum low-temperature Createc STM with a base pressure $<10^{-10}$ mbar. Electrochemically etched tungsten tips were prepared by \textit{in situ} sputtering and preparation on Au, and were likely Au terminated.
All $I$ vs $V$ curves were numerically differentiated to generate the $\frac{dI}{dV}$ spectra. No other treatment of the data was performed prior to the \textit{k}-means analysis.

\section*{Visualization of \textit{k}-means functionality}\label{sup:other_kmeans}

\begin{figure*}[!h]
    \centering
    \includegraphics[width=14.0cm]{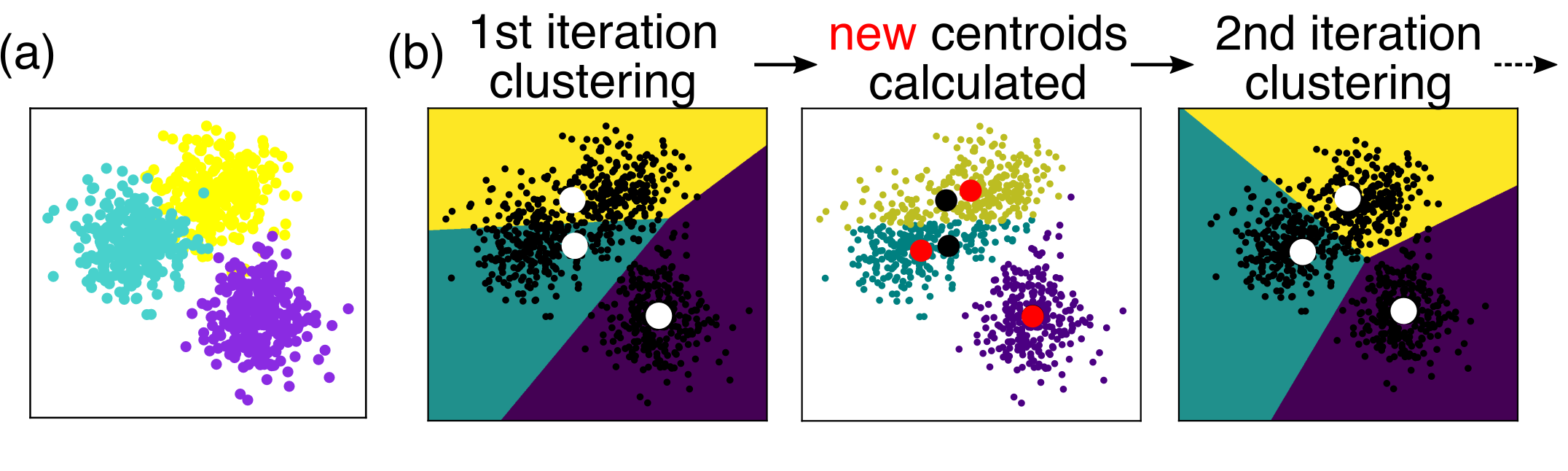}
    \caption{Depiction of \textit{k}-means algorithm for two-dimensional toy data with 3 clusters. (a) the toy dataset of three groups of Gaussian-spread data points coloured according to identity. (b) demonstration of the step-by-step process of the \textit{k}-means algorithm treating the toy dataset.}
    \label{supfig:1}
\end{figure*}

Figure \ref{supfig:1} provides a visual demonstration of the steps of the \textit{k}-means algorithm using a toy dataset for simplicity. The toy data used is three Gaussian `blobs' -- groups of points centered at arbitrary coordinates with added Gaussian-distributed noise. 
\textit{k}-means was then applied to the toy dataset with $k=3$. Unlike the `k-means++' method used in our analysis, Figure \ref{supfig:1} uses simple \textit{k}-means which picks initial centroids from the data points randomly rather than enforcing some centroid disparity. Next, all data points were assigned to one of the three centroids by minimizing Euclidean distance. We can see from the first panel in Figure \ref{supfig:1} (b) that these initial centroids happen to distinguish one of the `blobs' well, while the remaining two are not appropriately clustered. 

After this first clustering of the data points to the centroids, the center of mass of each cluster is calculated. This center of mass value is then set as the next iteration's centroids. The second panel of Figure \ref{supfig:1} (b) shows how this significantly shifts two of the centroids towards better separating the `blobs'. Finally, the data points are reclustered according to the new centroids. This process iterates until a preset number of iterations has occurred, or until the centroids change minimally between subsequent iterations according to a predefined threshold. 

\begin{figure}[!h]
    \centering
    \includegraphics[width=8.5cm]{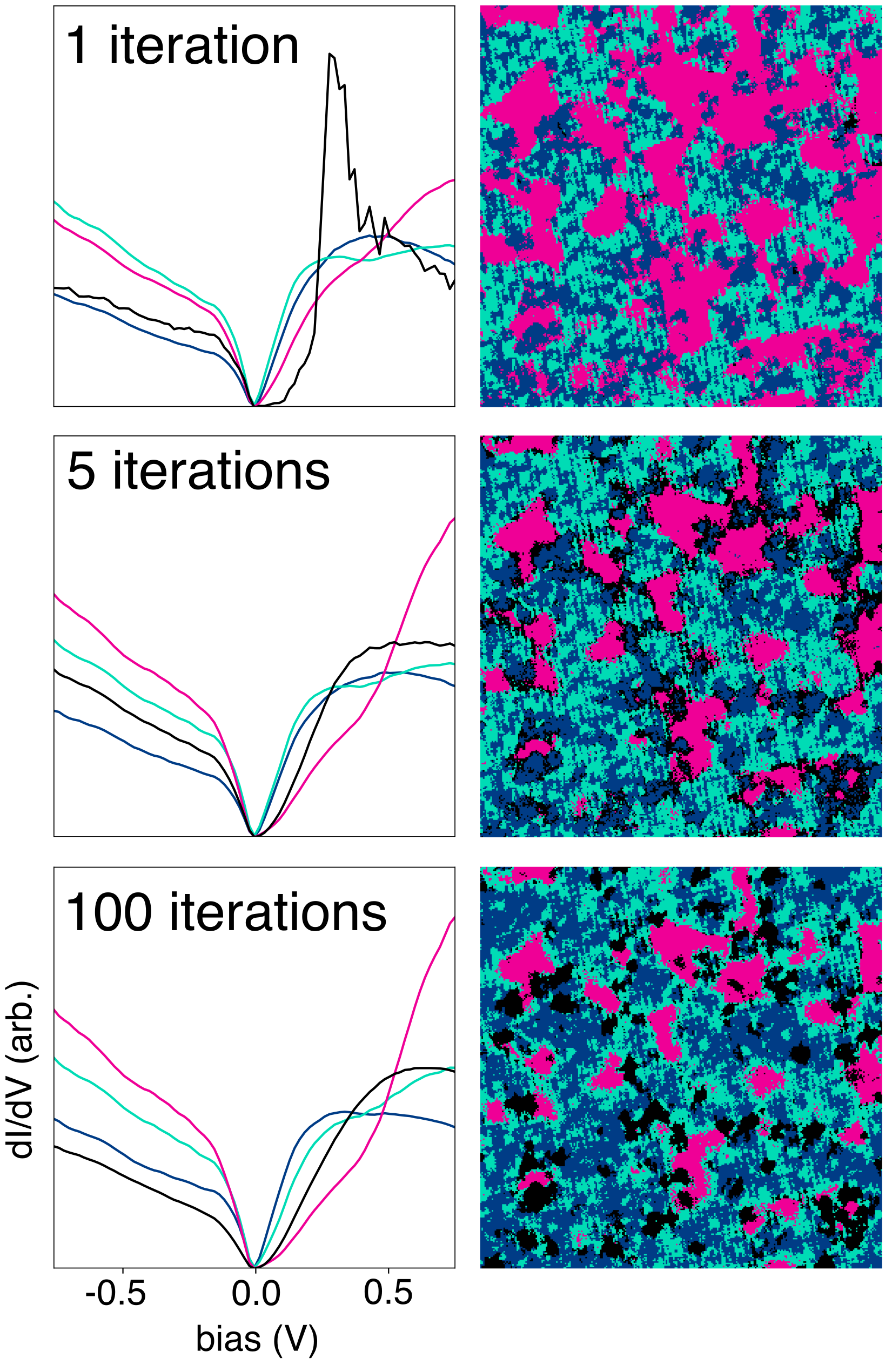}
    \caption{Depiction of the \textit{k}-means algorithm iterating on a Sr$_2$Ir$_{0.95}$Rh$_{0.05}$O$_4$ STM/STS grid map using a random initialization of the centroids.}
    \label{supfig:2}
\end{figure}

Figure \ref{supfig:2} demonstrates this same progression of the \textit{k}-means algorithm but for the dataset analyzed in this letter. Note that here we use simple \textit{k}-means, as opposed to the `k-means++' method, so that the change in the centroids are more visually obvious. After 100 iterations, the centroids and spatial cluster distribution appear nearly identical to the results in main text Figure 1 (c) and (d). 

\begin{figure}[!h]
    \centering
    \includegraphics[width=11cm]{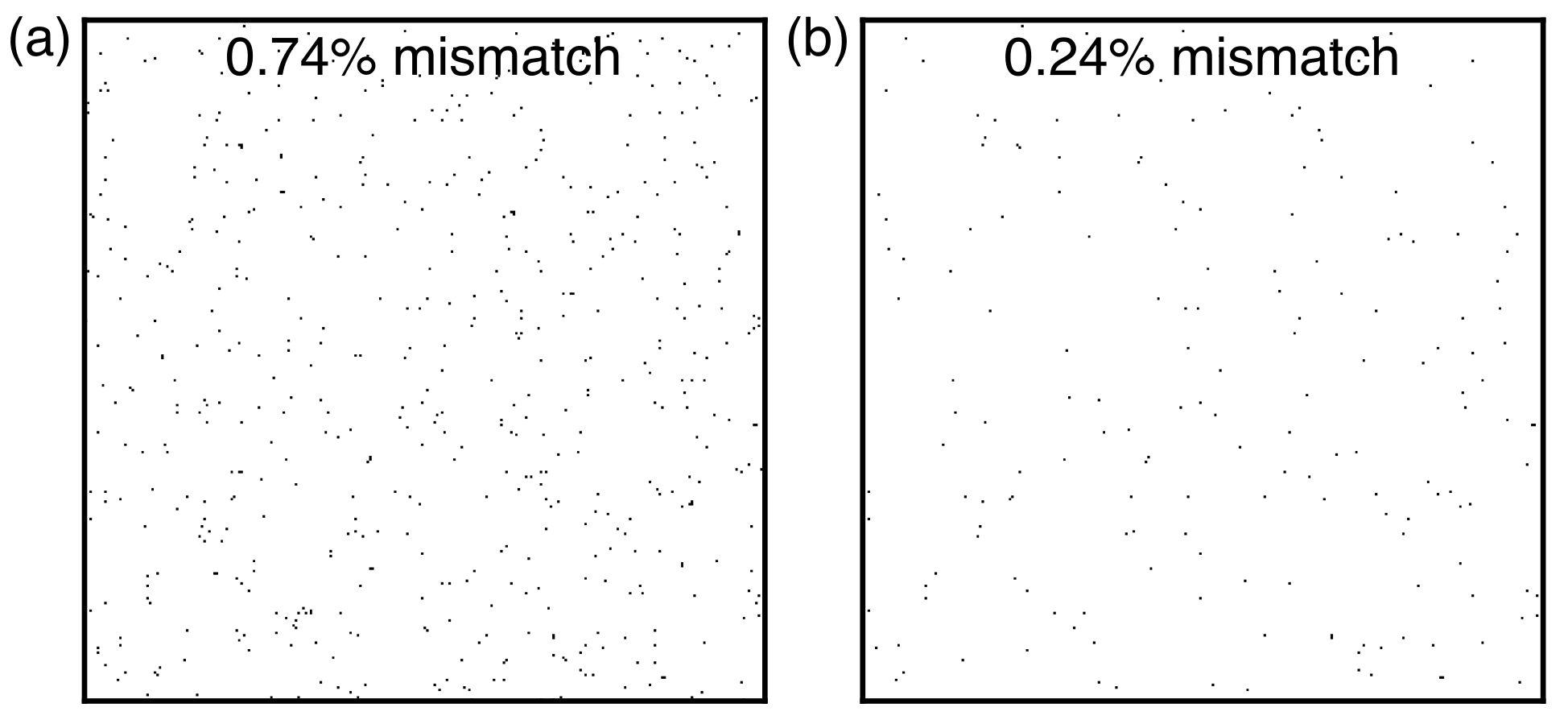}
    \caption{Pixels of cluster identity mismatch resulting from the initialization method `k-means++' versus (a) manual centroid initialization using the example spectra from main text Figure 1 (a), and (b) random centroid initialization. \text{k}-means calculated on the Sr$_2$Ir$_{0.95}$Rh$_{0.05}$O$_4$ STM/STS grid map discussed in the main text. The best out of 10 random initializations was chosen to avoid a local minimum. }
    \label{supfig:3}
\end{figure}

Figure \ref{supfig:3} demonstrates the difference in the \textit{k}-means results based off the different initialization methods available using the scikit-learn implementation of \textit{k}-means. The initialization methods are `k-means++', which chooses more disparate initial centroids, manually selecting the initial centroids, or randomly selecting the initial centroids. The \textit{k}-means results in the main body of this letter was done with `k-means++'. The difference in the cluster assignment between the `k-means++' and manual initial centroid methods was 0.74\% of all point spectra. The difference in the cluster assignment between the `k-means++' and random initial centroid methods was 0.24\% of all point spectra. Given theses mismatches are a small percentage of all spectra and that they appear randomly distributed in Figure \ref{supfig:3}, we used only `k-means++' as it was the fastest to compute.

\section*{STM/STS grid map of Sr$_2$Ir$_{0.82}$Rh$_{0.18}$O$_4$ measured at 4.5 K.}

Similar to the Sr$_2$Ir$_{0.95}$Rh$_{0.05}$O$_4$ STM/STS grid map examined in the main body of this letter, an STM/STS grid map of Sr$_2$Ir$_{0.82}$Rh$_{0.18}$O$_4$ measured at 4.5 K was analyzed using \textit{k}-means. 

\begin{figure}[!h]
    \centering
    \includegraphics[width=10cm]{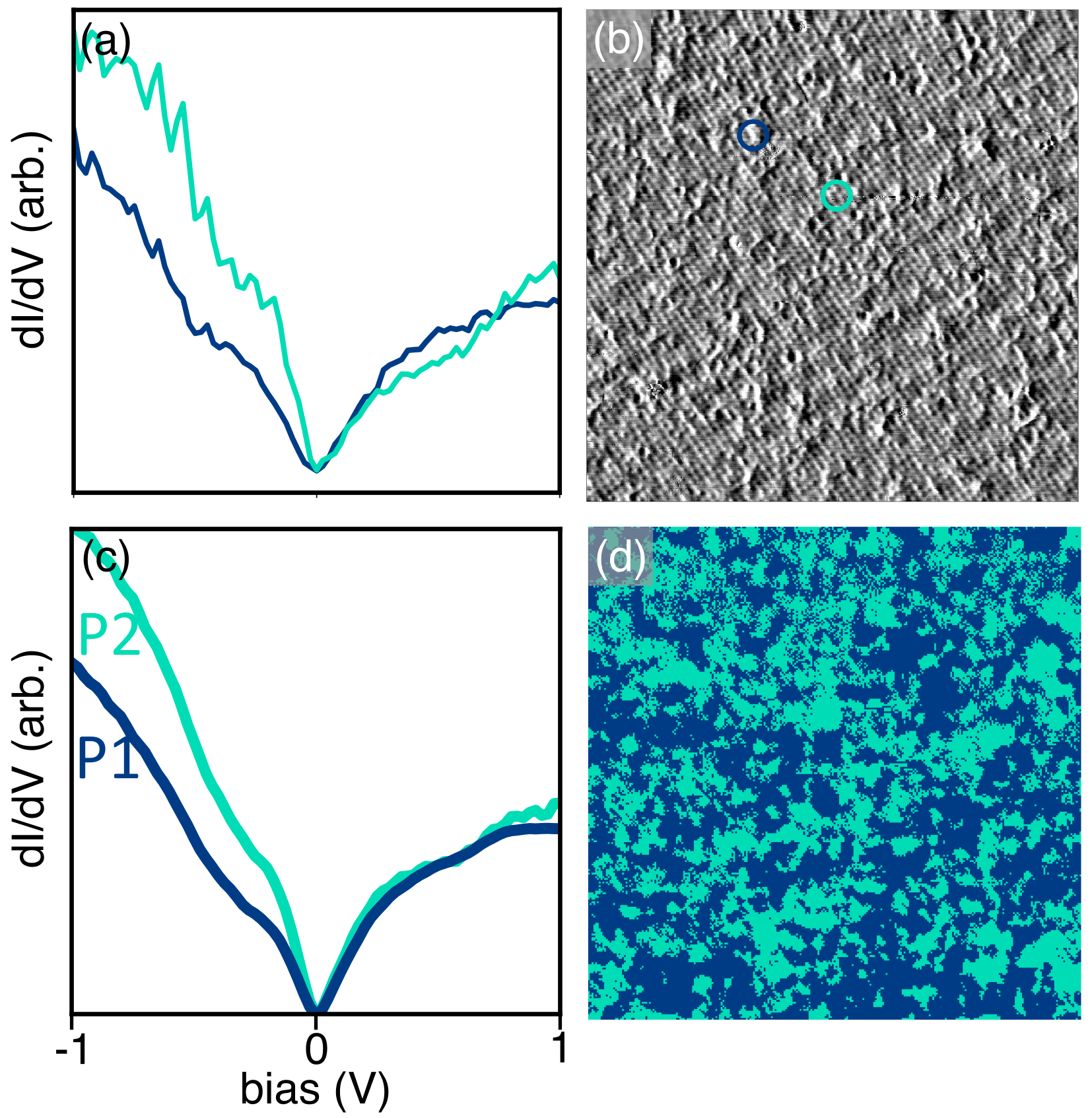}
    \caption{(a) Example spectra picked out by eye from (b) a 27 nm x 27 nm STM topography image of 18\% Rh-doped Sr$_2$IrO$_{4}$ at 4.5 K, taken at 800 mV and 400 pA. Positions of the spectra from (a) are circled. (c) -- (d) results from \textit{k}-means applied to the same grid map with $k=2$. (c) centroids of the two clusters, named according to their spectral shape, and (d) spatial distribution of the two clusters.}
    \label{supfig:4}
\end{figure}

\begin{figure}[!h]
    \centering
    \includegraphics[width=10cm]{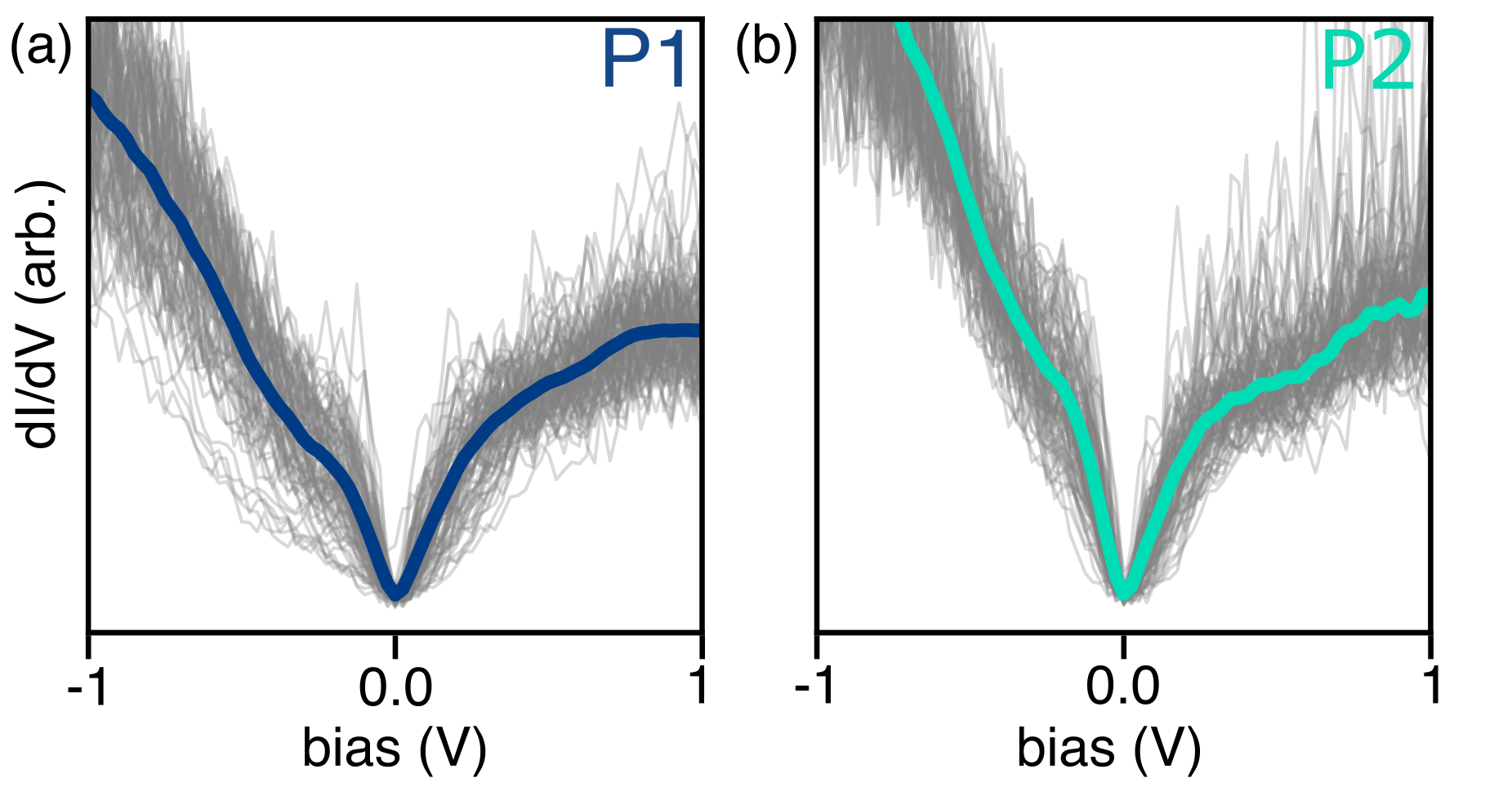}
    \caption{Centroids and randomly sampled spectra from the 18\% Rh-doped Sr$_2$IrO$_{4}$ at 4.5 K dataset belonging to (a) cluster P1 and (b) cluster P2. }
    \label{supfig:5}
\end{figure}

It is thought that this 18\% Rh sample is in the hidden order phase\cite{Zhao2016}, making it unclear if there would be any remaining Mott-like behaviour. Through both \textit{k}-means and manual searching, no qualitatively Mott phase areas were found in the data set. Additionally, increasing $k$ above 2 did not result in a centroid shaped like a defect cluster: additional centroids all appeared qualitatively pseudogap-like. The distribution of spectra belonging to each cluster in Figure \ref{supfig:5} is consistent enough that we are confident that we are not missing significant areas of unique electronic behaviour.

Fitting of equation 2 from the main text to this dataset was attempted, but failed within physically reasonable values of variable parameters. We suggest this is likely due to no Mott gap shaped spectra being present.

\section*{STM/STS grid map of Sr$_2$Ir$_{0.82}$Rh$_{0.18}$O$_4$ measured at 77 K}

An STM/STS grid map of the same Sr$_2$Ir$_{0.82}$Rh$_{0.18}$O$_4$ sample was measured at 77 K. Like the previous measurement at 4.5 K, it was again unclear if there would be any remaining Mott-like behaviour.\cite{Zhao2016} Through both k-means and manual searching, no qualitatively Mott phase areas were found in the data set. Like the measurement at 4.5 K, increasing $k$ above 2 did not result in a centroid shaped like a defect cluster: all centroids appeared qualitatively pseudogap-like.

\begin{figure}[H]
    \centering
    \includegraphics[width=10cm]{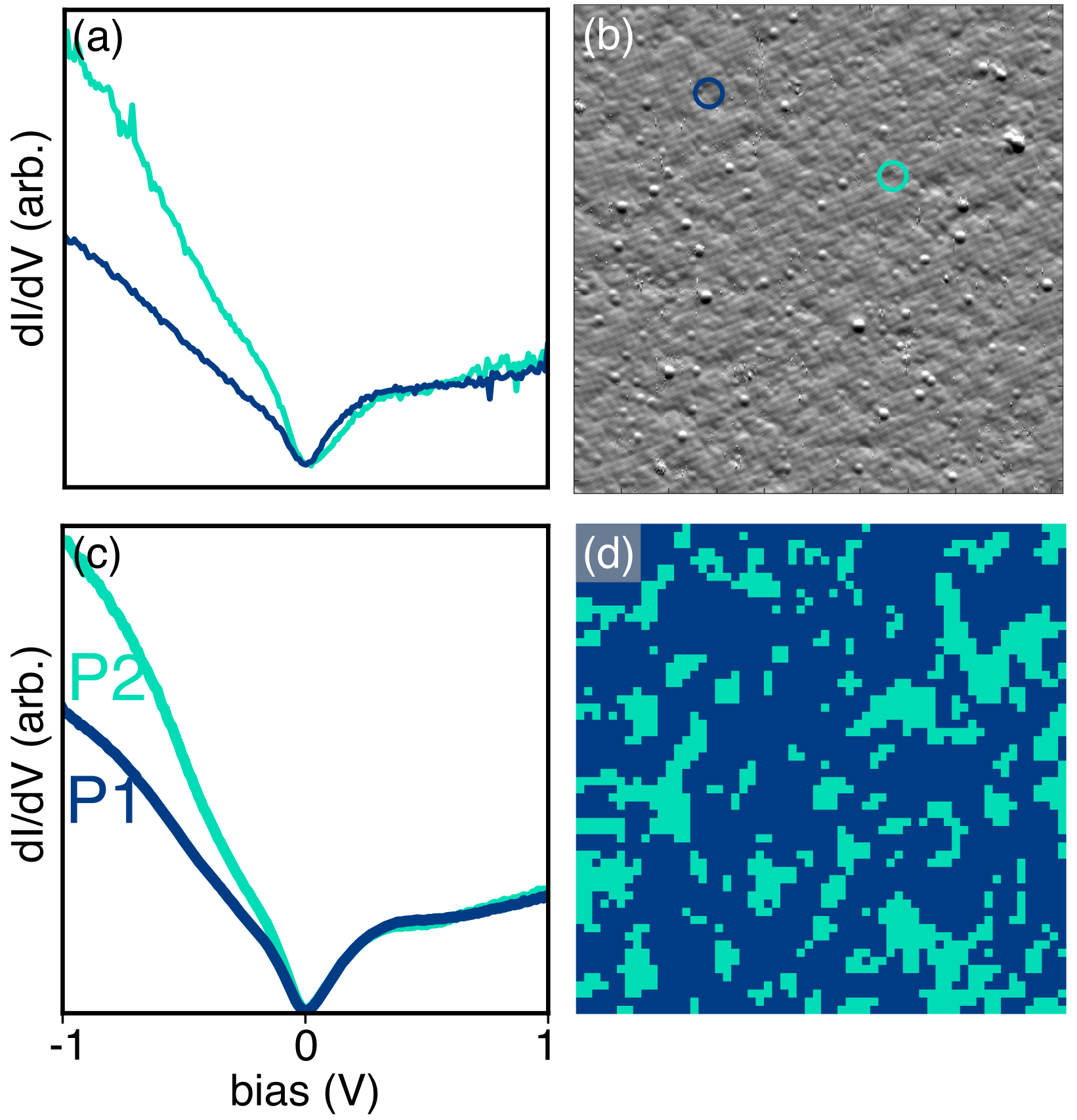}
    \caption{(a) Example spectra picked out by eye from (b) a 30 nm x 30 nm STM topography image of 18\% Rh-doped Sr$_2$IrO$_{4}$ at 77 K, taken at 1 V and 300 pA. Positions of the spectra from (a) are circled. (c) -- (d) results from \textit{k}-means applied to the same grid map with $k=2$. (c) centroids of the two clusters, named according to their spectral shape, and (d) spatial distribution of the two clusters.}
    \label{supfig:6}
\end{figure}

\begin{figure}[H]
    \centering
    \includegraphics[width=10cm]{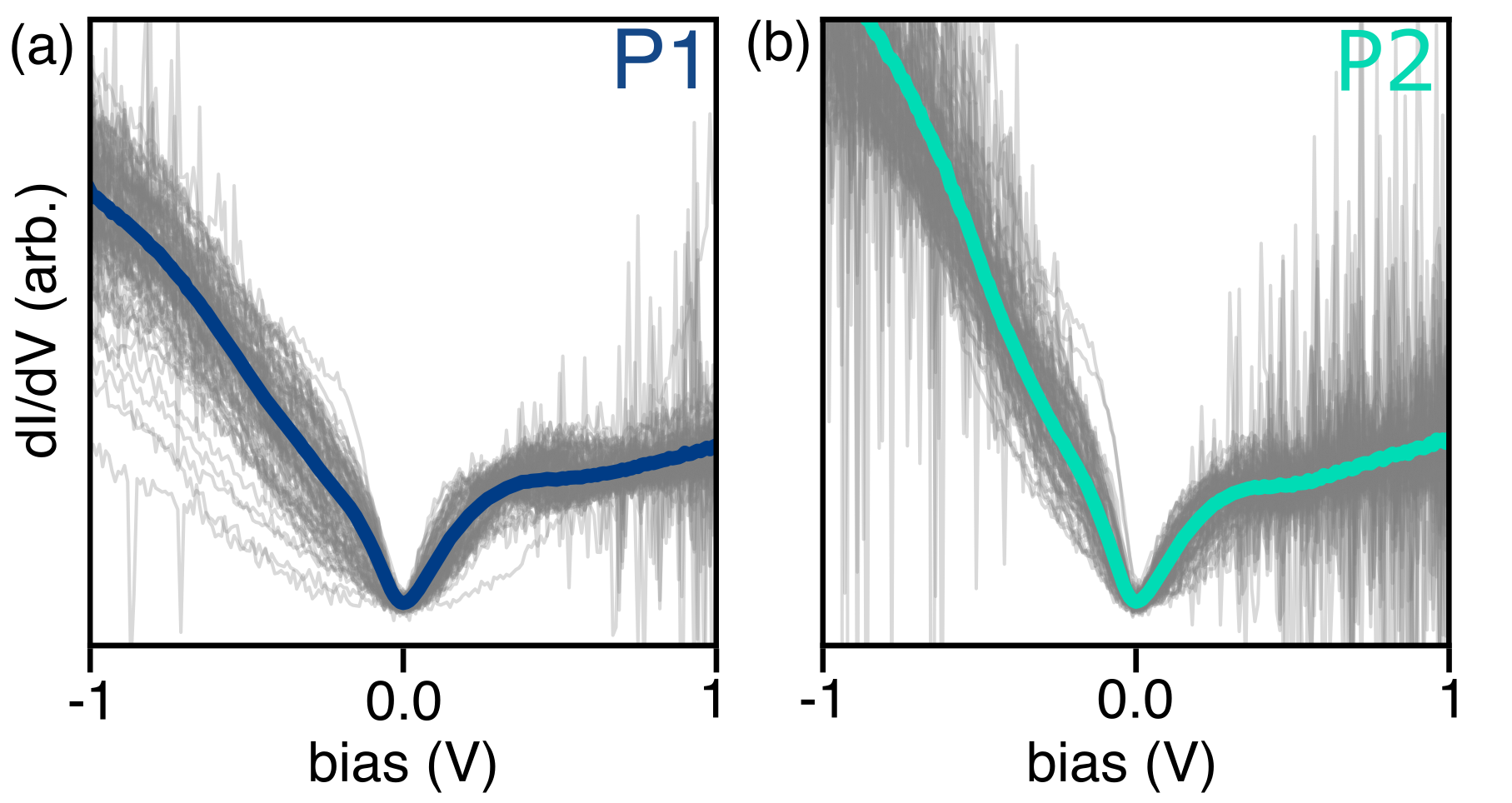}
    \caption{Centroids and randomly sampled spectra from the 18\% Rh-doped Sr$_2$IrO$_{4}$ at 77 K dataset belonging to (a) cluster P1 and (b) cluster P2.}
    \label{supfig:7}
\end{figure}

The distribution of spectra belonging to each cluster in Figure \ref{supfig:7} was deemed consistent enough that we were not missing significant areas of unique electronic behaviour.

Fitting of equation 2 from the main text to this dataset was attempted, but failed within physically reasonable values of variable parameters. Again, we suggest this is likely due to no Mott gap shaped spectra being present.

\bibliographystyle{unsrt}
\bibliography{k-means}